\documentclass[preprint,showpacs,preprintnumbers,amsmath,amssymb]{revtex4}

\usepackage{graphicx}
\usepackage{epstopdf}
\newcommand{\ba}{\hbox{\large a}}
\newcommand{\be}{\hbox{\large e}}

\begin{document}

\title{Analytic Approximation of the Tavis-Cummings Ground State via
Projected States \\}

\author{Octavio Casta\~nos, Eduardo
Nahmad-Achar, Ram\'on L\'opez-Pe\~na, and Jorge G. Hirsch}

\affiliation{Instituto de Ciencias Nucleares, Universidad Nacional
Aut\'onoma de M\'exico, Apdo. Postal 70-543 M\'exico 04510 D.F.\\ \\}

%\vspace{0.2in}

\begin{abstract}

We show that an excellent approximation to the exact quantum solution of the ground state of the Tavis-Cummings model is obtained by means of a semi-classical projected state. This state has an analytical form in terms of the model parameters and, in contrast to the exact quantum state, it allows for an analytical calculation of the expectation values of field and matter observables, entanglement entropy between field and matter, squeezing parameter, and population probability distributions. The fidelity between this projected state and the exact quantum ground state is very close to 1, except for the region of classical phase transitions. We compare the analytical results with those of the exact solution obtained through the direct Hamiltonian diagonalization as a function of the atomic separation energy and the matter-field coupling.
\end{abstract}


\maketitle

\section{Introduction}

The progress in the technology of trapped-atom lasers and the cavity QED experiments has resumed the interest of researchers in the basic models that describe the interaction between quantized radiation and atoms. The Dicke Model~\cite{dicke} (DM) describes the interaction of a quantized radiation field with a sample of $N$ two-level atoms located within a distance smaller than the wavelength of the radiation.  Dicke realized that under certain conditions a gas of radiating molecules shows a collective behavior called {\it superradiance}.  This phenomenon was observed experimentally in optically pumped HF gas~\cite{exp1}. The simplest case $N=1$ in the rotating-wave approximation is known as the Jaynes-Cummings model (JCM)~\cite{jc1963}. Many theoretical predictions of this latter model, such as the existence of collapse and revivals in the Rabi oscillations~\cite{eberly}, the formation of macroscopic quantum states, or measures of entanglement associated with spin-squeezed states, have been confirmed, and many experimental studies of Rydberg atoms with very large principal quantum number within single-mode cavities have been observed~\cite{manko}. It is well known that the JCM has served as a guide to understand several quantum optics phenomena. While this considers the rotating wave approximation in order to discard the non-conserving energy terms, the model is widely applicable: in 1991 it was proposed~\cite{crisp} that by using circularly polarized light together with atomic selection rules the non-conserving energy terms can be eliminated; additionally, the diamagnetic term can be taken into account by properly changing the frequency of the field mode as a function of the coupling interaction strength: the renormalization of the field mode frequency can be understood by making a Bogoliubov transformation on the field part of the Dicke Hamiltonian. This implies that, at least in some cases, the results of the JCM can be valid beyond the rotating wave approximation.

The JCM, which might well be called the standard model of quantum optics, has been studied intensively for more than forty years.  An experimental example is provided by Aoki et al.~\cite{aoki}.  When there are many atoms interacting with a single mode quantized radiation field of one and the same cavity, the exact solution can be obtained by means of the so-called {\it Tavis-Cummings} model~\cite{tavis} (TCM), and its recent generalizations~\cite{duke}. This model predicts the collective $N$-atom interaction strength to be $\gamma_{N}=\sqrt{N}\,g_{i}$, where $g_{i}$ is the dipole coupling strength of each individual atom $i$. The TCM also has been considered to describe cavity QED with a Bose-Einstein condensate~\cite{brennecke}.  Exact solutions of the generalized Tavis-Cummings model have been found by means of quantum inverse methods, an algebraic procedure based on finding sets of solutions to the Bethe equations~\cite{bogoliubov}.  Other algebraic methods solve the eigenvalue problem of the TCM using polynomially deformed algebras~\cite{vadeiko} where analytical expressions may be found up to third order in a specific perturbation theory.

Since the presence of superradiant phase transitions in the DM were established~\cite{lieb}, there have been several contributions simplifying the original computation and others that have found also the presence of phase transitions in generalized Dicke models~\cite{hioe1, hioe2, comer}. In particular, one of the contributions showed the existence of phase transitions in a Dicke Hamiltonian which includes the counter-rotating atom-field interaction, although the necessary condition between the coupling parameter and the field mode frequency was modified~\cite{comer}. 

In 1975 it was shown that the superradiant phase transition was due to the absence of the diamagnetic term in the Hamiltonian describing the N two-level atoms interacting with a one mode electromagnetic field~\cite{wod1}. It established, through the Thomas-Reiche-Kuhn sum rule, that the phase transition cannot be reached because it would place contradictory bounds to the parameters of the model. Other authors~\cite{knight,bir1,kaz1} have established that gauge invariance requires the presence of the diamagnetic and counter-rotating terms; moreover, the sum rule can be derived from charge-current conservation. For this reason, a criterion was seeked to establish the validity of Dicke models and determine whether a superradiant phase transition can occur. It was shown that two-level atoms which make electric dipole transitions cannot exhibit superradiant phase transitions, although their result does not rule out phase transitions based on magnetic dipole interactions. In the last decade, the work of Crisp was extended to many two-level atoms and it was concluded that in the strong coupling regime, in the thermodynamical limit, it is not possible to have superradiant phase transitions~\cite{liberty}.

It is however known that, for a one electron atom interacting with a non-quantized electromagnetic field (that is, in the semiclassical approximation), a unitary transformation of the form $U(\vec{r},t) = \exp\left(\frac{i\, e}{\hbar \, c}\, \vec{r}\cdot \vec{A}(0,t)\right)$ can be performed to eliminate the terms depending on the electromagnetic vector potential, giving rise to an electric dipole-field interaction. The same procedure can be applied to the quantized electromagnetic field, the difference being that we are dealing with operators and there is an extra term, due to the commutation relations, which represents a dipole-dipole interaction~\cite{milonni}. In 2001, similar arguments on the interaction between radiation and molecular dipole moments were given~\cite{siva}, where they use a gauge invariant formulation of the molecular electric dipole-photon interactions which allows for superradiant phase transitions as in the Hepp-Lieb formulation. They show that the dipole-field interaction is strictly linear in the electric field, with an extra self-interaction term. Thus, there are no quadratic terms, and therefore a physical system displaying superradiant phase transitions should exist. The difference with previous gauge invariant formulations is due to the time dependence of the unitary transformation. 

The considerations above render likely the physical reality of systems presenting superradiant phase transitions, and being described by a Hamiltonian of the type used to describe linear dipole-photon interactions. In particular, there are physical systems where interactions between bosonic degrees of freedom with a collective spin~\cite{milburn, tobias} are relevant.  Then, the recently presented method~\cite{scrip} which allows a simple and elegant determination of the stability properties and phase transitions for a finite number of particles, is of importance.

In this work we obtain an excellent approximation to the superradiant states of the ground state of the Tavis-Cummings Hamiltonian, which admits analytical expressions for both, field and matter observables, including the entanglement entropy between field and matter, the squeezing parameter, and the population probability distribution. The fitness between this approximation and the exact quantum solution is measured through the evaluation of the fidelity parameter\~cite{wooters}.

Section II below establishes the Tavis-Cummings Hamiltonian in terms of the constant of motion, and shows that the variational tensorial product of coherent states has difficulties in describing, for example, the photon number fluctuations. It is argued that restoring the symmetry by projecting the variational state to a given value of the constant of motion, a much better (and analytical) approximation is obtained.  This projected state is justified and built. In Section III, by means of the overlap of the projected state, the calculation of the expectation values for field and matter observables is done analytically for an arbitrary value of the constant of motion. The entanglement entropy between the matter and field, the squeezing parameter of the state and the probability distributions of photons and atoms are also calculated. In Section IV the selection of the appropriate value of the constant of motion $\lambda$ is discussed. Section V compares the results of Section III with the corresponding ones for the exact quantum state, and Section VI draws some conclusions.

\section{Projected State}

The Tavis-Cummings model Hamiltonian for $N$ identical $2$-level systems (e.g., atoms) immersed in an electromagnetic field, is given by
	\begin{equation}
		H_{TCM}=\omega_{F}\,\ba^{\dagger}\ba+\tilde\omega_{A}\,J_{z}
		+\frac{\gamma_N}{\sqrt{N}}\left(\ba^{\dagger}\,J_{-}
		+\ba\,J_{+}\right)\ ,
		\label{HTCM}
	\end{equation}
where $\omega_F$ is the field frequency, $\tilde\omega_{A}$ the atomic energy-level difference, and $\gamma_N$ the dipole coupling. The operators $\ba$, $\ba^{\dagger}$, denote the one-mode annihilation and creation photon operators, $J_{z}$ the atomic relative population operator, and $J_{\pm}$ the atomic transition operators.

It is immediate that this Hamiltonian commutes with the operator 
	\begin{equation}
		\Lambda = \ba^{\dagger}\ba+J_{z}\ .
		\label{constant-motion}
	\end{equation}
It is then convenient to rewrite it
by introducing a detunning parameter $\tilde\Delta = \omega_F - \tilde\omega_{A}$, and by dividing it by 
$\omega_{F}$ (which can be thought of as the natural unit of frequency) and by the total number of particles, having in this way an intensive Hamiltonian operator
	\begin{equation}
		H=\frac{1}{N}\,\Lambda -  \frac{\Delta}{N}\,J_{z}
		+\frac{\gamma}{\sqrt{N}\,N}\left(\ba^{\dagger}\,J_{-}
		+\ba\,J_{+}\right)\ ,
		\label{jaynes-cummings}
	\end{equation}
where $\Delta= 1-\frac{\tilde\omega_{A}}{\omega_F} \equiv 1-\omega_A$ and $\gamma= \frac{\gamma_N}{\omega_F}$.

\begin{figure}[h]
\scalebox{0.75}{\includegraphics{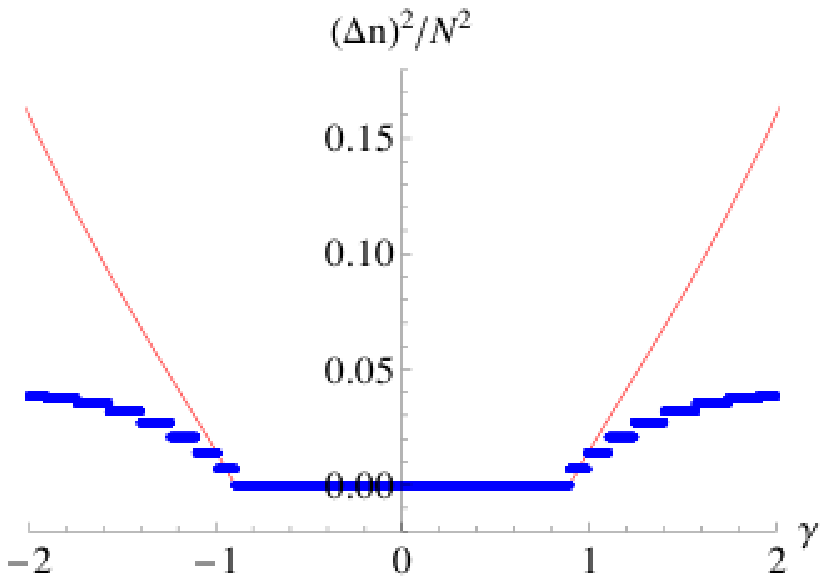}}
\qquad
\scalebox{0.75}{\includegraphics{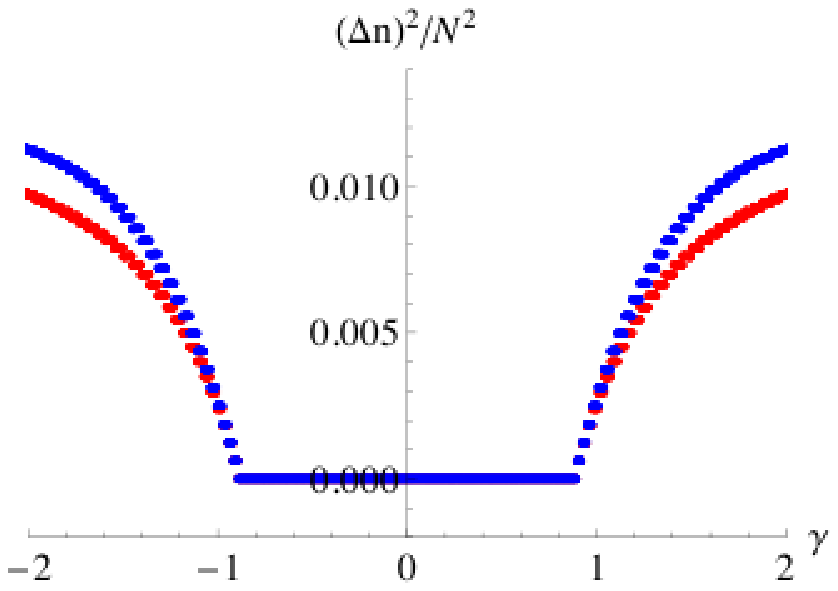}}
\caption{\label{delta-n2}
Squared fluctuation of the photon number operator $\hat{n}$, for the ground state, as a function of the interaction strength $\gamma$, for a detuning parameter $\Delta=0.2$ and $N=20$ atoms. At left, the continuous line shows the coherent state approximation to the exact solution (discrete line). The differences are large even when normalized by the number of atoms. At right the lower curve shows the approximation we can obtain using a projected state, as explained in this paper. Note the different scales for both plots.}
\end{figure}

Following the variational procedure indicated in~\cite{scrip}, it was found that the direct product
of coherent states: Heisenberg-Weyl for the photon
part $\vert\alpha\rangle$ \cite{meystre,manko} and $SU(2)$ or spin for the matter $\vert\zeta\rangle$ \cite{gilmore1972}, i.e., $\vert\alpha, \,\zeta\rangle=\vert\alpha\rangle \otimes\vert\zeta\rangle$, is a good approximation to the exact quantum solution of the Tavis-Cummings Hamiltonian, in spite of the fact that the overlap of the two states is small. This approximation describes very well the expectation values of
the atomic observables and most of the field observables. However, particular difficulties are found in the description of the photon number fluctuations $(\Delta\hat{n})^{2}$, as shown in the left of Fig.(\ref{delta-n2}), and for several observables when the detuning parameter $\Delta$ takes values greater than $1$.

%Figura delta-n2

The first of these problems arises because the variational state has contributions from all the eigenvalues $\lambda$ of the operator $\Lambda$, violating its conservation symmetry. It is then natural to propose as ground state a state with the symmetry restored by projecting the variational state to a given value of $\lambda$, namely, the one that minimizes the classical energy obtained from the variational procedure and approximated to the closest integer or half-integer according to whether $N$ is even or odd, respectively. The exact procedure is presented in Section IV below.

By projecting the state to one (as yet arbitrary) value of $\lambda$ we obtain
	\begin{equation}
	\vert \psi \rangle = 
	\begin{cases}
		\vert 0\rangle\otimes\vert j,\,-j\rangle \ ,
		&\text{$\omega_{A}>\gamma^{2}$}\ ;\\
		\sum_{\nu=\max[0,\,\lambda-j]}^{\lambda+j}\,
		\binom{2j}{j+\lambda-\nu}^{1/2}\,\be^{-2\,i\,\nu\,\phi}
		\,\frac{\zeta^{\nu}}{\sqrt{\nu!}}\,\vert\nu
		\rangle\otimes\vert j,\,\lambda-\nu\rangle\ ,
		&\text{$\left|\omega_{A}\right|\leq \gamma^{2}$}\ ;\\
		\vert 0\rangle\otimes\vert j,\,j\rangle\ ,
		&\text{$\omega_{A}<-\gamma^{2}$}\ .
		\label{projected-state}
	\end{cases}
	\end{equation}
In this expression we have used $j=\frac{N}{2}$, and for $\left|\omega_{A}\right|\leq \gamma^{2}$ the state is not normalized, with
	\begin{equation}
		\zeta=-\frac{\sqrt{N}\,\gamma}{2} \,\,\left(1
		+\frac{\omega_{A}}{\gamma^{2}}\right)\ .
	\end{equation}
It has the advantage that it only contains one eigenvalue of the constant of
motion, it is an analytic solution, reproduces well the matter and field observables (cf. Fig~(\ref{delta-n2}) at right and note the scale difference with the figure at left), and the overlap with the exact quantum solution is very close to one ({\it vide infra}). 
	
The conditions on the parameters $\omega_A$ and $\gamma$ in Eq.(\ref{projected-state}) are associated to the existence of a separatrix in the model which defines the phase transitions in the semiclassical solution~\cite{scrip}.

To build the projected state Eq.(\ref{projected-state}) we use the semiclassical procedure, which consists of the calculation of the expectation value of the Hamiltonian with respect to the tensorial product of coherent states $\vert\alpha\rangle \otimes\vert\chi\rangle$, the determination of the minima, and the use of the catastrophe formalism~\cite{gilmore3} to find the stability properties~\cite{scrip}. We find that the critical points for a minimum are given by  
	\begin{eqnarray*}
		q_{c}&=&-\sqrt{j}\,\gamma\,\sin\theta_{c}\,\cos\phi_{c}\ ,
	\label{quadratureq}\\
		p_{c}&=&\phantom{-}\sqrt{j}\,\gamma\,\sin\theta_{c}\,\sin\phi_{c}\ ,
	\end{eqnarray*}
where we have written $\alpha= \frac{1}{\sqrt{2}}(q + i p)$ in terms of the expectation values of the quadratures of the field, and $\chi=e^{i \phi} \tan{(\frac{\theta}{2})}$, with $(\theta,\phi)$ denoting a point in the unit Bloch sphere; at their critical values $\theta_{c},\ \phi_c,$ we have~\cite{scrip}
	\begin{equation}
		\hbox{minima:} \left\{
		\begin{array}{llll}
		\theta_{c}=0\,,&E_{0}=-\frac{N\,\omega_{A}}{2}\,,&\lambda_{c}=-j \, ,&\hbox{ for }\omega_{A}>			\gamma^2\\
		\theta_{c}=\pi\,,&E_{0}=\phantom{-}\frac{N\,\omega_{A}}{2}\,,&\lambda_{c}=\phantom{-}j \, ,
		&\hbox{ for }\omega_{A}<-\gamma^2\\
		\theta_{c}=\arccos\left(\frac{\omega_{A}}{\gamma^{2}}\right)\,,
		&E_{0}=-\frac{N(\omega_{A}^{2}+\gamma^{4})}{4\,\gamma^{2}}\,,&
		\lambda_{c}=j\,\frac{-\omega_{A}\,\left(\omega_{A}+2\right)+\gamma^{4}}{2\,\gamma^{2}} \, , &
		\hbox{ for }\left|\omega_{A}\right|<\gamma^{2} 
		\end{array}\right.
		\label{critical}
	\end{equation}
This expression shows the minima critical points, the energy $E_0$, the constant of motion $\lambda_c$, and the conditions in the parameter space to guarantee that they constitute an energy minimum.  The energy surface of the Tavis-Cummings model is $\phi$-unstable, for which reason $\phi_c$ can be taken arbitrarily.

The expression for the trial state $\vert \alpha \rangle \otimes \vert \chi \rangle$ that minimizes the energy surface takes the following form~\cite{scrip}:
	\begin{eqnarray}
		\hbox{{\it North Pole} ($\omega_{A}>\gamma^{2}$):} \qquad
		\vert\psi_{np}\rangle&=&\vert 0\rangle\otimes\vert j,\,-j\rangle
		\label{polo-norte}\\
		\hbox{{\it South Pole} ($\omega_{A}<-\gamma^{2}$):} \qquad 	
		\vert\psi_{sp}\rangle&=&\vert 0\rangle\otimes\vert j,\, j\rangle
		\label{polo-sur}\\
		\hbox{{\it Parallels} ($\left|\omega_{A}\right|<\gamma^{2}$):} \qquad 	
		\vert\psi_{par} \rangle&=&\sum_{m=-j}^{+j}\ \sum_{\nu=0}^{+\infty}\,
		A_{m,\,\nu} \vert\nu\rangle\otimes\vert j,\,m\rangle
	\end{eqnarray}
where we have defined the expansion coefficients
	\begin{eqnarray*}
		A_{m,\,\nu} &=&\binom{2j}{j+m}^{1/2}\, \exp\left\{-\frac{j\, \gamma^{2}}{4}\left(1-\frac{
		\omega_{A}^{2}}{\gamma^{4}}\right) + i\,(j+m-\nu)\,\phi \,\right\}
		\nonumber \\
		&&\times \frac{\left(-\sqrt{2j}
		\,\gamma\right)^{\nu}}{\sqrt{\nu!}}
		\left(\frac{1}{2}+\frac{\omega_{A}}{2\,\gamma^{2}}\right)^{(j-m+\nu)/2}\,
		\left(\frac{1}{2}-\frac{\omega_{A}}{2\,\gamma^{2}}\right)^{(j+m+\nu)/2}\, .
	\end{eqnarray*}
One usually would perform this sum by selecting a value of $\nu$, making the sum over
$m$, and then proceeding with the following value of $\nu$, etc., until we
reach some type of convergence. In the TCM model $\lambda=m+\nu$ is a
conserved quantity. By replacing $m$ by $\lambda-\nu$, we can write~\cite{scrip}
	\begin{equation}
		\vert\psi_{par} \rangle=\sum_{\lambda=-j}^{+\infty}
		\sum_{\phantom{of}\nu=\max(0,\lambda-j)}^{\lambda+j}
		A_{\lambda-\nu,\,\nu}\ \vert\nu\rangle \otimes \vert j,\, \lambda-\nu\rangle\ .
		\label{paralelo}
	\end{equation}

The eigenstates of the North Pole~(\ref{polo-norte}), South Pole~(\ref{polo-sur}),
and Parallel region~(\ref{paralelo}), for a given $\lambda$, justify the projected
state established in (\ref{projected-state}) with the unnormalized state for
$|\omega_{A}|\le\gamma^{2}$ defined by
	\begin{equation}
		\vert\zeta;\,j,\lambda\} \equiv \vert\zeta\} =
		\sum_{\nu=\max[0,\,\lambda-j]}^{
		\lambda+j}\,
		\binom{2j}{j+\lambda-\nu}^{1/2}\,\be^{-2\,i\,\nu\,\phi}
		\,\frac{\zeta^{\nu}}{\sqrt{\nu!}}\,\vert\nu\rangle\otimes
		\vert j,\,\lambda-\nu\rangle\ ,
		\label{projected-par}
	\end{equation}
where we have retained only terms depending on the number of photons $\nu$ from the
coefficient $A_{\lambda-\nu,\,\nu}$.

\section{Expectation Values of Field and Matter Observables}

The effect of the transformation generated by the constant of motion
$\exp[{i\varphi\Lambda}]$ in the plane of quadratures of
the electromagnetic field is a counterclockwise rotation by an angle $\varphi$, while for the matter observables $J_x$ and $J_y$ it is a clockwise rotation by an angle $\varphi$ along the $z$-axis. This
transformation then leaves invariant each term in the TCM Hamiltonian. This gives our trial states their $\phi$-invariance. Calculating for any value of $\phi$ allows one to recover the result for any other value of $\phi$ through the rotation above. Without loss of generality, then, we choose $\phi=0$ in what follows.

To determine the expectation values of the observables of the system, in the Parallels region, it is useful to calculate the overlap
	\begin{equation}
		\{\zeta^{\prime}\vert\zeta\} = \sum_{\nu=\max[
		0,\,\lambda-j]}^{\lambda+j}\,
		\binom{2j}{j+\lambda-\nu}\,\frac{\left(\zeta^{\prime}\,\zeta
		\right)^{\nu}}{\nu!}\ .
	\end{equation}
Considering this expression as a truncated expansion of the
Hypergeometric Confluent function we obtain
	\begin{equation}
		\big\{\zeta^{\prime}\vert\zeta\big\}=\left\{
		\begin{array}{ll}
		L_{j+\lambda}^{j-\lambda}(-\zeta\,\zeta^{\prime})
		&,\quad\left|\lambda\right|\le j\\
		\frac{(2j)!}{(j+\lambda)!}\,\left(\zeta\,
		\zeta^{\prime}\right)^{\lambda-j}\,
		L_{2j}^{\lambda-j}(-\zeta\,\zeta^{\prime})
		&,\quad\lambda\ge j
		\end{array}\right.
		\label{traslape}
	\end{equation}
where $L_{n}^{\alpha}(x)$ denotes tha associated Laguerre
polynomials, which are defined as~\cite{andrews2000}
	\begin{eqnarray*}
		L_{n}^{\alpha}(x)&=&\frac{x^{-\alpha}\,e^{x}}{n!}\,
		\frac{d^{n}\phantom{x}}{dx^{n}}\left(e^{-x}\,
		x^{n+\alpha}\right)\ ,\quad n\ge 0\ ,\\
		\left(x\,\frac{d\phantom{x}}{dx}\right)\,
		L_{n}^{\alpha}(x)&=&n\,L_{n}^{\alpha}(x)-(n+\alpha)\,
		L_{n-1}^{\alpha}(x)\ ,\quad n\ge 1\ .
	\end{eqnarray*}

From expression~(\ref{projected-state}) for the projected state, it
is immediate that the probability of finding $\nu$ photons depends of
the relative values of $\omega_{A}$ and $\gamma$, i.e.,
	\begin{equation}
		{\cal P}_{\nu}=\frac{1}{\big\{\zeta\vert\zeta\big\}}\,
		\frac{\zeta^{2\nu}}{\nu!}\,\binom{2j}{j+\lambda-\nu}\ ,
		\label{probnu}
	\end{equation}
for $\left|\omega_{a}\right|\le\gamma^{2}$, while
${\cal P}_{\nu}=\delta_{\nu,\,0}$ outside that region. The probability of finding $n_{e}$
excited atoms can be obtained from the previous expression by replacing
$\nu\rightarrow\lambda+j-n_{e}$.

Defining $\eta\equiv\zeta^{2}$, and the overlap of the projected state by
$Y\equiv\big\{\zeta\vert\zeta\big\}$, we can calculate
the expectation value of the photon number
operator $\hat{n}=\ba^{\dagger}\ba$ and its corresponding fluctuations squared $(\Delta \hat{n})^{2}$ :
	\begin{eqnarray}
		\langle \hat{n}\rangle&=&\sum_{\nu=\max[0,\,
		\lambda-j]}^{\lambda+j}\,\nu\,{\cal P}_{\nu}
		=\eta\,\frac{d\phantom{\eta}}{d\eta}
		\,\ln Y\ ,
		\label{esperado-n} \\
		(\Delta \hat{n})^{2}&=&\left(\eta\,\frac{d\phantom{\eta}}{d\eta}
		\right)^{2}\,\ln Y\ .
		\label{fluctuations-n}
	\end{eqnarray}
Note that for $\lambda=-j$ the overlap is $Y=1$.  On the other hand, for $\lambda=j$ the overlap is $Y=L_{2j}^{0}(-\zeta^2)$, which only equals $1$ when $\omega_A=-\gamma^2$.  At the North and South Poles, therefore, the expectation values for $\hat{n}$ and $(\Delta \hat{n})^{2}$ can be obtained by assuming $Y=1$ there and the expressions (\ref{esperado-n}), (\ref{fluctuations-n}) to be valid for all values of the parameters $\gamma$ and $\omega_A$.  Analytic expressions for $\langle \hat{n}\rangle$ and $(\Delta \hat{n})^{2}$ in the Parallels region are given in the Appendix.

The expectation values of other matter and field observables can be determined in terms of the equations above.

The quadrature components of the electromagnetic
field  are given by
	\[
		\hat{q}=\frac{1}{\sqrt{2}}\left(\ba^{\dagger}+\ba\right)\
		,\qquad \hat{p}=\frac{1}{\sqrt{2}\,i}\left(\ba^{\dagger}
		-\ba\right)\ ;
	\]
as they change the value of $\lambda$ for any state, their expectation values with respect to the projected state
are equal to zero. Their corresponding fluctuations are therefore
	\begin{equation}
		\langle\hat{q}^{2}\rangle=\langle\hat{p}^{2}\rangle
		=\langle\hat{n}\rangle+\frac{1}{2}\ .
	\end{equation}
For the expectation values of matter observables we have $\langle J_{x}\rangle = \langle J_{y}\rangle = 0$ for the same reason as above, and others take the form
	\begin{eqnarray}
		\langle J_{z}\rangle&=&\lambda-\langle
		\hat{n}\rangle\ ,\\
		\left(\Delta J_{z}\right)^{2}&=&\left(\Delta
		\hat{n}\right)^{2}\ ,\\
		\langle J_{x}^{2}\rangle&=&
		\langle J_{y}^{2}\rangle
		=\frac{1}{2}\,j(j+1)-\frac{1}{2}\,
		\left(\lambda-\langle\hat{n}\rangle\right)^{2}
	\end{eqnarray}
Finally, the expectation values for the transition
operators $\ba^{\dagger}J_{-}$ and $\ba\,J_{+}$ are given by
	\begin{equation}
		\langle\ba^{\dagger}\,J_{-}\rangle
		=\langle\ba\,J_{+}\rangle=\zeta\,\left(j
		+\lambda-\langle\hat{n}\rangle\right)\ .
	\end{equation}

\section{Determination of the constant of the motion}

The expectation value of the Hamiltonian with respect to the projected state for an arbitrary value of $\lambda$ is
	\begin{equation}
		\langle\hat{H}\rangle = \frac{\{\zeta\vert\hat{H}\vert\zeta\}}{\{\zeta\vert\zeta\}} =
		 \frac{\lambda}{2\,j}\left(1
		-\Delta\right)+\frac{2\,\gamma}{(2\,j)^{3/2}}\,\left(j
		+\lambda\right)\,\zeta
		+\frac{1}{2\,j}\,\left[\Delta-\frac{2\,\gamma}{\sqrt{2\,j}}
		\,\zeta\right]\langle\hat{n}\rangle
	\end{equation}
When $\lambda = -j$ we have $\langle\hat{n}\rangle = 0$ and this expression simplifies to $\langle\hat{H}\rangle=-\frac{1}{2}\left(1-\Delta\right)$.
By substituting the expression of the expectation value of $\hat{n}$ given in Eq.~(\ref{n-ph}) of the Appendix we obtain the energy surface
	\begin{equation}
		{\cal H}(\zeta)
		=\frac{\lambda+j\,\Delta}{2\,j}-\left[\Delta
		-\frac{2\,\gamma}{\sqrt{2\,j}}\,\zeta\right]
		\left\{\begin{array}{ll}
		L_{j+\lambda-1}^{j-\lambda}(-\zeta^{2})
		\Big/L_{j+\lambda}^{j-\lambda}(-\zeta^{2})&
		,\quad -j+1\le\lambda\le j\\
		\frac{\lambda+j}{2\,j}\,
		L_{2j-1}^{\lambda-j}(-\zeta^{2})
		\Big/L_{2j}^{\lambda-j}(-\zeta^{2})&
		,\quad\lambda\ge j
		\end{array}\right.
		\label{e-surface}
	\end{equation}
and ${\cal H}(\zeta)=-\frac{1}{2}\left(1-\Delta\right)$ for $\lambda = -j$.

\begin{figure}[h]
\scalebox{1.0}{\includegraphics{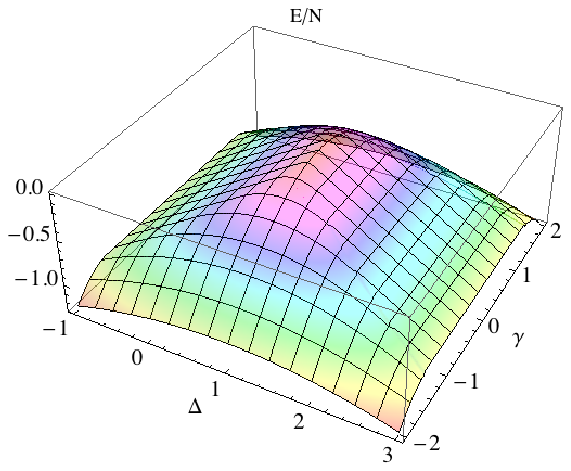}}\qquad\quad
\scalebox{1.0}{\includegraphics{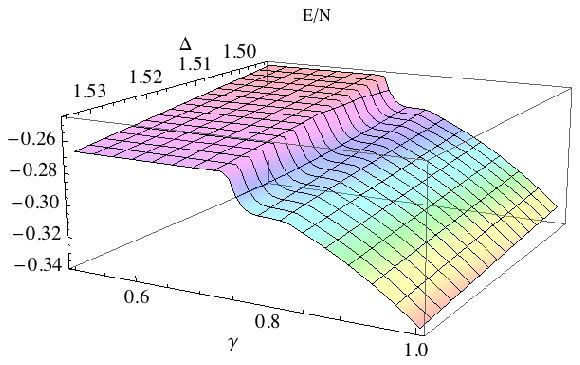}}
\caption{\label{energy3D}
Energy surface of the ground state used to calculate the value $\lambda$ of the constant of motion, for $N=20$ atoms. The zoom at right shows phase transitions when $\Delta$ is greater than 1 (see text for details).}
\end{figure}

\begin{figure}[h]
\scalebox{1.0}{\includegraphics{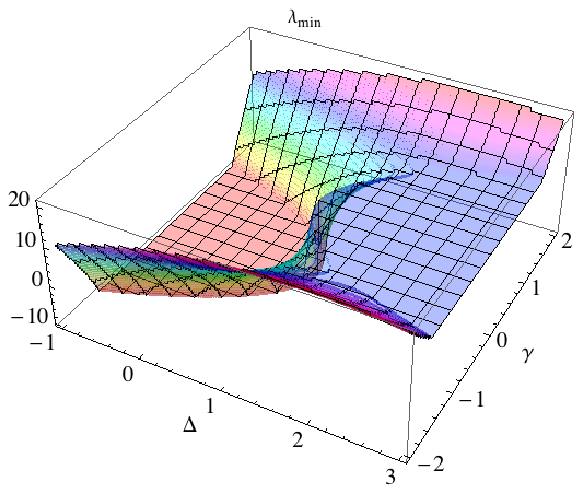}}\qquad\qquad
\scalebox{0.8}{\includegraphics{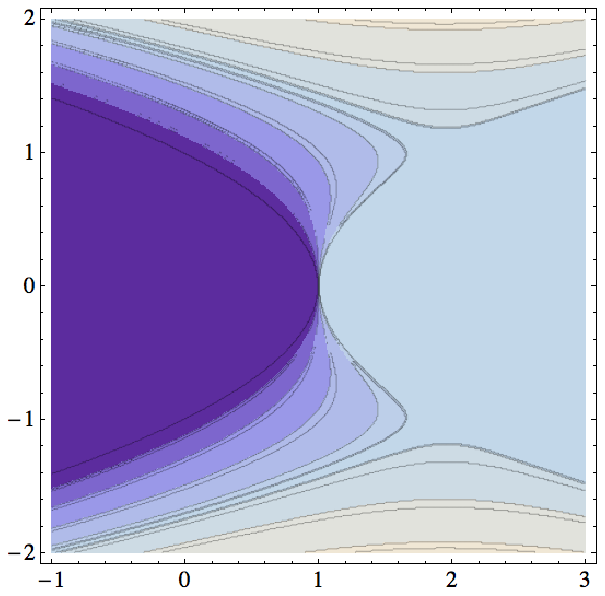}}
\caption{\label{lambda3D}
Constant of motion $\lambda_{min}$ determined by minimizing the expectation value of the Hamiltonian with respect to the projected state, as a function of the interaction strength $\gamma$ and the detuning parameter $\Delta$, for $N=20$ atoms. Contour levels are shown at right.}
\end{figure}

We calculate this expression as a function of $\Delta$ and $\gamma$, and for all possible values of $\lambda$ with $\Delta \in [-2,4]$ and $\gamma \in [-5,5]$, which allows us to choose the value of $\lambda$ for which the energy is minimum and onto which the coherent state is to be projected. By noting that the classical value $\lambda_c$ of the constant of motion, given in Ec.~(\ref{critical}), when rounded to the nearest integer or half-integer, never differs by more than one unit from its exact quantum counterpart, the minimizing procedure simplifies to testing for $\vert \lambda-\lambda_c \vert\ \le 1$, and approximating $\lambda$ to the nearest integer (if $N$ is even) or half-integer (if $N$ is odd). The minimum energy so obtained is plotted in Fig.(\ref{energy3D}), while the resulting constant of motion and its contour levels are shown in Fig.(\ref{lambda3D}).

%Figura Energy 3D

While similar, the behaviour for $\Delta > 1$ appears to be different than that for $\Delta \le 1$. The energy surface shows a fictitious phase transition near $\gamma \approx 0.7$ for $\Delta > 1$ (cf. right of Fig.(\ref{energy3D})) that does not exist for $\Delta \le 1$.  In fact, the regime $\Delta > 1$ appears only when $\omega_{A} < 0$; one may visualize physical situations where this may occur: if the atoms are immersed in an external magnetic field and the energy levels appear as a Zeeman splitting of spectral lines, one may think of continuously tuning the field intensity until a sign reversal is obtained, thus interchanging the excited and base levels. It will be seen, however, that the exact quantum solution does not show this transition (cf. Fig.(\ref{energy}) below). The case $\Delta = 1$ (or equivalently $\omega_{A} = 0$) is, nevertheless, a special and boundary case: the Hamiltonian in Ec.~(\ref{HTCM}) simplifies considerably, the atomic levels are degenerate, the quantum energy spectrum (see Section V) is also highly degenerate, and the poles (both North and South) contract to a single point.  In fact (cf. Ec.~(\ref{critical})), the North Pole is no longer a minimum critical region, but a saddle point.

%Figura lambda 3D

\begin{figure}[h]
\scalebox{1.0}{\includegraphics{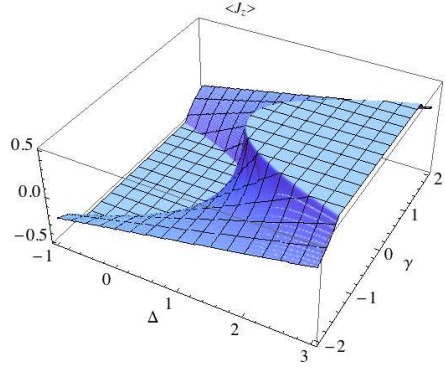}}
\caption{\label{jz3D}
Expectation value of the population inversion operator $J_{z}$, for the variational ground state,
as a function of the interaction strength $\gamma$ and the detuning parameter $\Delta$, for $N=20$
atoms.}
\end{figure}

From the expressions derived for the observables in Section III, their expectation values can also be plotted as functions of the interaction strength $\gamma$ and the detuning parameter $\Delta$. It will be seen that their behavior will be inherited from that of $\lambda_{min}$ itself. As an example, we show $\langle J_z\rangle$ in Fig.(\ref{jz3D}).

%Figura Jz3D

\begin{figure}[h]
\scalebox{1.0}{\includegraphics{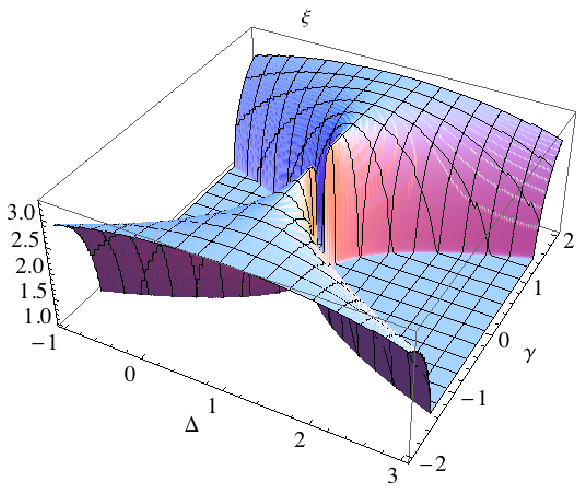}}
\qquad
\scalebox{1.0}{\includegraphics{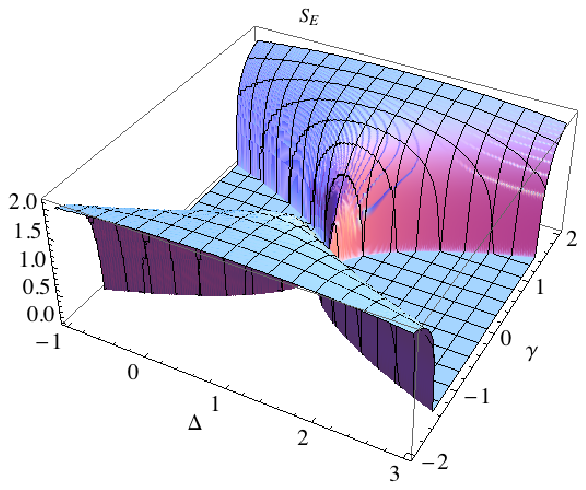}}
\caption{\label{Comp-Se3D}
Squeezing parameter $\xi$ for the matter (left), and entanglement entropy $S_E$ between field and matter (right), for the variational ground state, as a function of the interaction strength $\gamma$ and the detuning parameter $\Delta$, for $N=20$ atoms.}
\end{figure}

There is no field squeezing since, from the quadrature components, $\langle\hat{q}\rangle = \langle\hat{p}\rangle = 0$ and $\langle\hat{q}^2\rangle = \langle\hat{n}\rangle + \frac{1}{2} = \langle\hat{p}^2\rangle$. However, for the matter squeezing coefficient we have~\cite{kitagawa}
	\begin{equation}
		\xi = \sqrt{\frac{2(\Delta J_\perp)^2}{j}}
		\label{squeezing}
	\end{equation}
where $J_\perp$ is a component of $\vec{J}$ transverse to $\langle \vec{J} \rangle$; as $\langle \vec{J} \rangle = \langle J_z \rangle \hat{e}_z$, we take $J_x$ to be this component, and show $\xi$ in Fig.(\ref{Comp-Se3D}) (left). $\xi$ gives us information of how good the trial projected state is approximating the exact solution of the eigenvalue problem of the TCM Hamiltonian, and this comparison will be made in Section V below.

%Figura Compression3D & Entropy3D

Finally, the entanglement entropy $S_E$ is zero for coherent states since these are expressed as product states. In~\cite{scrip}, in order to calculate a useful expression for $S_E$, we first traced over the field and then again over one of the matter modes. When a projected state is used, it is no longer a product state and we may trace over the matter (or equivalently over the field) to obtain
	\begin{equation}
		S_E = - \sum_{\nu} {\cal P}_{\nu} \log{{\cal P}_{\nu}}
		\label{se}
	\end{equation}
where ${\cal P}_{\nu}$ is already normalized, as given in Ec.~(\ref{probnu}). This is shown in Fig.(\ref{Comp-Se3D}) (right). Its behavior is very close to that found in~\cite{scrip}.

\section{Projected Coherent State vs. Exact Solution}

The exact solution for the ground state in the Tavis-Cummings model was presented in~\cite{scrip}, where the natural basis $\vert\nu\rangle\otimes\vert j,\,\lambda-\nu\rangle$ was used (natural because the Hamiltonian has $\Lambda$ as a constant of motion):
	\begin{equation}
		\vert\psi_{gs}\rangle=\sum_{\nu=\max[0,\lambda-j]}^{\lambda+j}\,c_{\nu}
		\,\vert\nu\rangle\otimes\vert j,\,\lambda-\nu\rangle\ ,
		\label{quantum-gs}
	\end{equation}
Analytical solutions for values of $\lambda$ up to $\lambda = - \frac{N}{2} + 4$	were also given, both in resonance ($\Delta = 0$)~\cite{buzek} and away from resonance ($\Delta \neq 0$)~\cite{scrip}. For greater values, numerical solutions were obtained.

\begin{figure}[h]
\scalebox{0.8}{\includegraphics{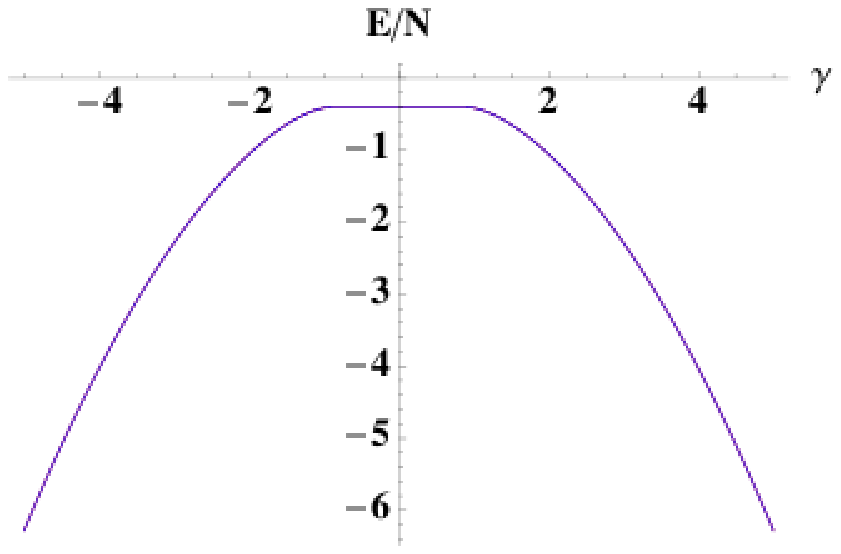}}
\qquad
\scalebox{0.8}{\includegraphics{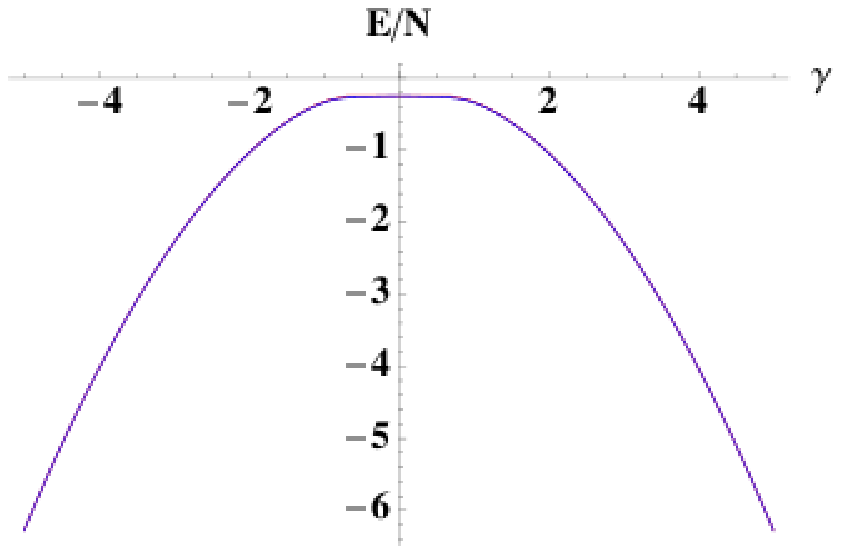}} \\
\vspace{0.2in}
\scalebox{0.8}{\includegraphics{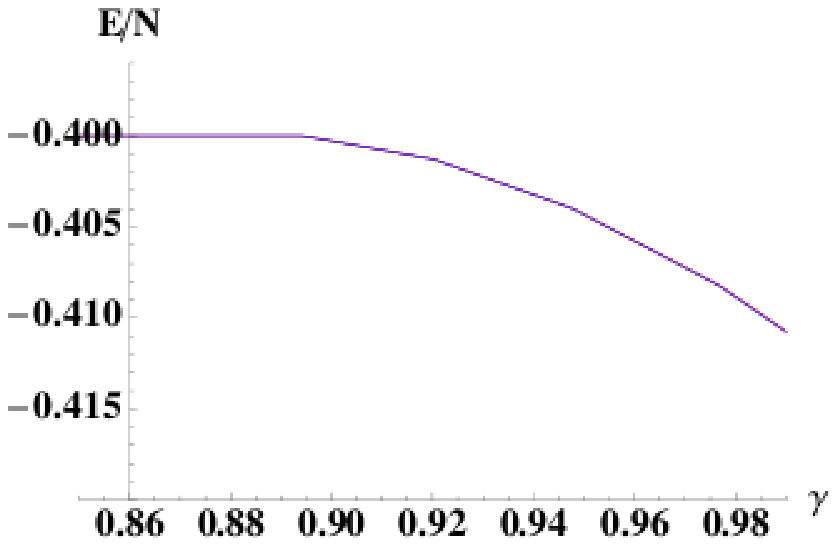}}
\qquad
\scalebox{0.8}{\includegraphics{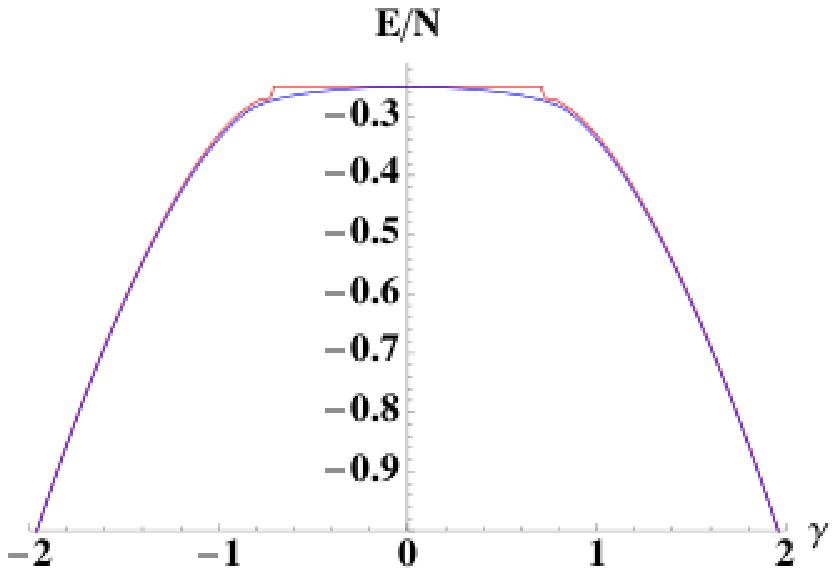}}
\caption{\label{energy}
Ground state energy per particle of the projected state solution compared with the exact quantum solution, as a function of the interaction strength $\gamma$ for $N=20$ atoms. The plots on the left correspond to a detuning parameter $\Delta=0.2$, while those on the right to a $\Delta=1.5$ (and therefore $\omega_A < 0$). The bottom plots are close-ups of those on top.  Note that both solutions lie on top of each other and are indistinguishable, except for the close-up for $\Delta=1.5$.}
\end{figure}

When the exact solution is compared with that obtained from the projected state, we find that the quantum phase transitions are exactly reproduced, as shown in Fig.(\ref{energy}), even for a large number of atoms ($N=20$) and when away from resonance. The straight lines in the lower left figure correspond to the energy of the ground state for different values of the constant of motion $\lambda$, starting from $-10$ ($0$ photons) and up to $-6$ ($4$ photons). In each case the eigenstate is a mixture of $0$ to $\lambda + \frac{N}{2}$ photons. The projected state approximation is stunningly close; so much so, that both graphs (quantum and projected state) practically lie on top of each other and are indistinguishable. Whereas the projected state has a constant value of the ground state energy inside the poles (North ($\vert 0\rangle\otimes\vert j,\,-j\rangle$), and South ($\vert 0\rangle\otimes\vert j,\,j\rangle$)), the exact quantum solution does not when $\Delta > 1\  (\omega_A < 0)$; the solution therefore is very precise for a detuning of $\Delta=0.2$, as shown in the left side of the figure, but fails slightly near the boundary of the poles for $\Delta=1.5$, as shown at lower right.

%Figura Energ'a

\begin{figure}[h]
\scalebox{1.0}{\includegraphics{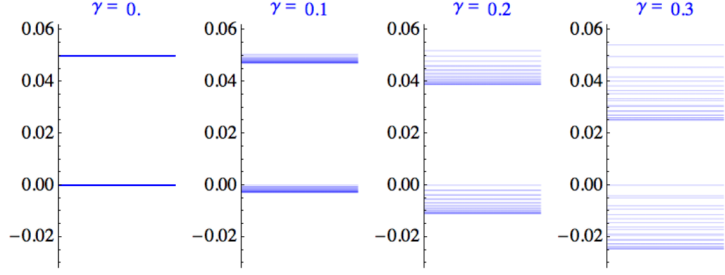}}
\caption{\label{energy_deg}
Energy spectrum for $\Delta = 1$. When $\gamma=0$ we have a $(2j+1)$-fold degeneracy (left plot shows the first $2$ such levels), which unfolds as soon as $\gamma$ is perturbed away from zero. Spectra for $\gamma=0,\ 0.1,\ 0.2$, and $0.3$ are shown.}
\end{figure}

The case $\Delta=1$ is a particular boundary, as mentioned in the previous section, since it simplifies the Hamiltonian~(\ref{jaynes-cummings}) making it independent of the inversion population operator $J_z$. In this case, if $\gamma=0$ also, we have a dense degeneracy of the Hamiltonian eigenvalues: the matter does not see the radiation field, so the degeneracy is that of the atoms themselves, i.e., $2j+1$. Any small deviation of $\gamma$ away from zero unfolds the energy levels, as shown in Fig.(\ref{energy_deg}). In this case our method for choosing the $\lambda$-value onto which to project fails, as $2j+1$ values yield the same energy. We shall, in what follows, choose values for $\Delta$ above and below $1$ to exemplify the behavior of observables.

%Figura E-degenerados

\begin{figure}[h]
\scalebox{0.7}{\includegraphics{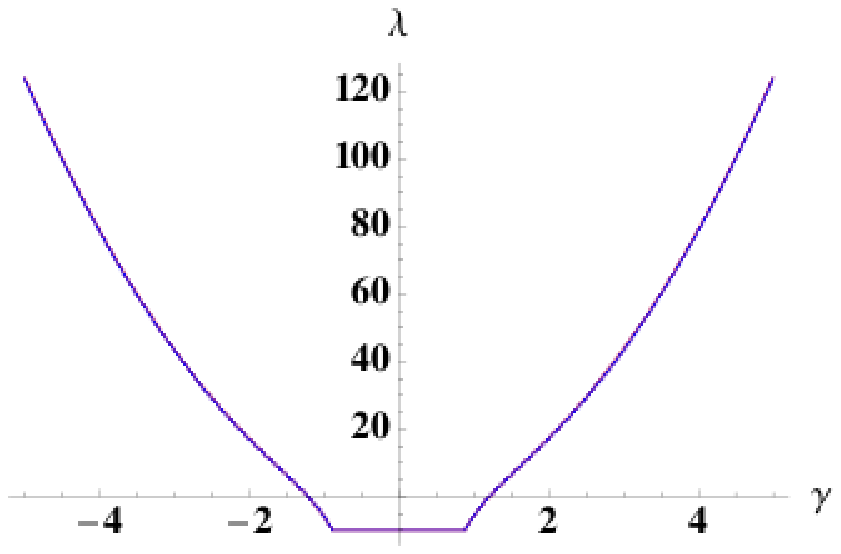}}
\qquad
\scalebox{0.7}{\includegraphics{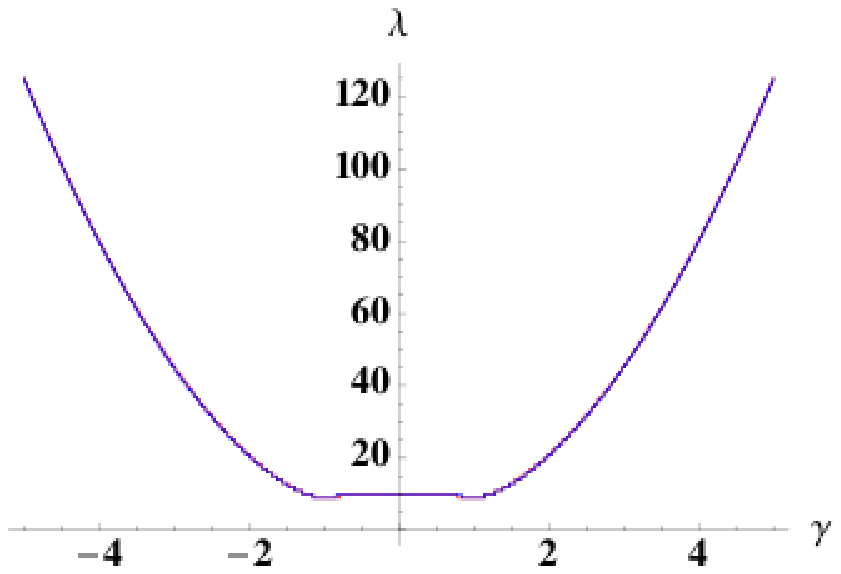}} \\
\scalebox{0.7}{\includegraphics{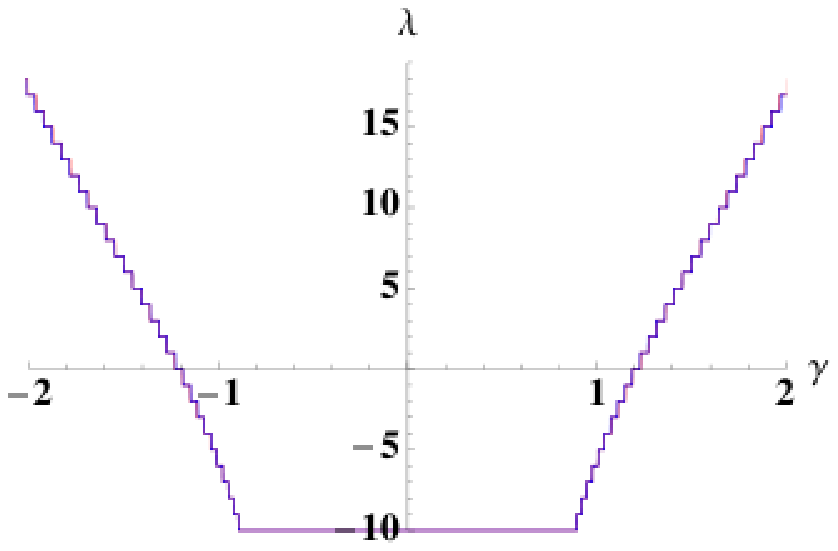}}
\qquad
\scalebox{0.7}{\includegraphics{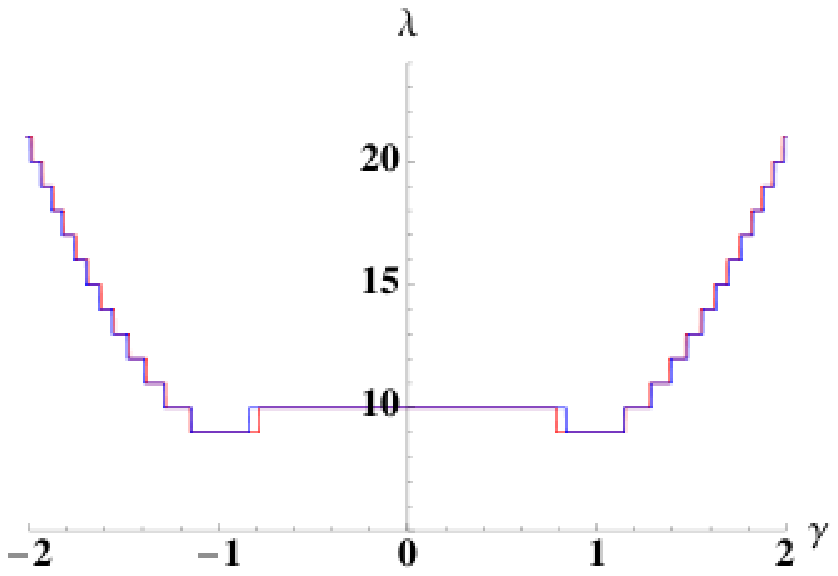}}
\caption{\label{lambda}
Constant of motion $\lambda_{min}$ of the projected state solution compared with the exact quantum solution, as a function of the interaction strength $\gamma$ for $N=20$ atoms. Note, once again, that both graphs practically lie on top of each other and are indistinguishable. The plots on the left correspond to a detuning parameter $\Delta=0.2$, while those on the right to a $\Delta=1.5$ (and therefore $\omega_A < 0$). The bottom plots are close-ups of those on top.}
\end{figure}

The constant of motion, $\lambda_{min}$ is shown in Fig.(\ref{lambda}) as a function of the interaction parameter $\gamma$ for both, the semiclassical projected and the quantum cases. Its discrete behavior is inherited by all the quantum observables of interest. The slope observed in the $\lambda$-steps is given precisely by the difference in energy between the absorbed photons and the atom's energy level separation. This is a consequence of the system being away from resonance ($\Delta \neq 0$).  When in resonance, the steps are horizontal. The plots at left show $\lambda$ for a detuning of $\Delta=0.2$, while those at right show it for $\Delta=1.5$. Note (lower right plot) that even when $\omega_A < 0$ the projected state solution tries to reproduce the non-constant behavior near the pole boundary. Otherwise the graphs are indistinguishable.

%Figura lambda

\begin{figure}[h]
\scalebox{0.9}{\includegraphics{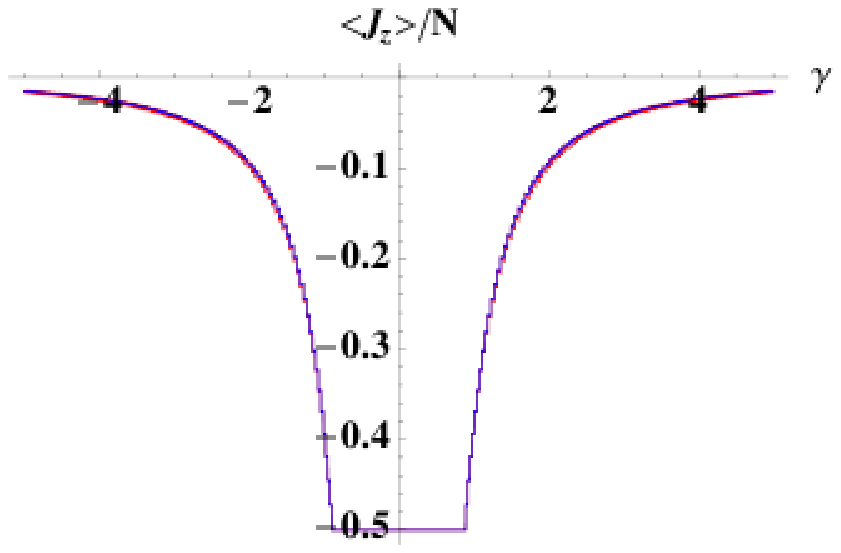}}
\qquad
\scalebox{0.9}{\includegraphics{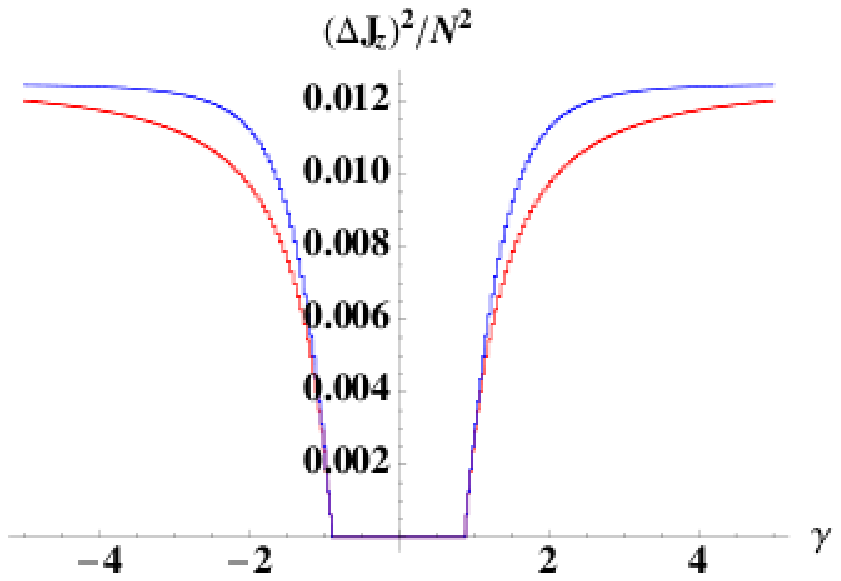}}
\caption{\label{Jz}
Expectation value for $J_z$ and its dispersion $(\Delta J_z)^2$ of the projected state solution compared with the exact quantum solution, as functions of the interaction strength $\gamma$ for $N=20$ atoms. Both graphs practically lie on top of each other (cf. the scale for $(\Delta J_z)^2$). The plots correspond to a detuning parameter $\Delta=0.2$.}
\end{figure}

The photon number fluctuations $(\Delta \hat{n})^{2}$, as noted in the Introduction, are very well resembled by the projected state solution to the TCM (cf. right of Fig.(\ref{delta-n2})). The differences are minute. The same trend is found for the dispersion in $J_z$, as is to be expected, since $\langle J_z \rangle = \lambda - \langle \hat{n} \rangle,\ \langle J_{z}^2 \rangle = \langle \hat{n}^2 \rangle - 2 \lambda \langle \hat{n} \rangle + \lambda^2$, and therefore $(\Delta J_z)^2 = (\Delta \hat{n})^2$. Both, the expectation value for $J_z$ and its dispersion are plotted in Fig.(\ref{Jz}) as functions of the interaction strength $\gamma$, slightly away from resonance ($\Delta = 0.2$). Note the scale in the ordinate axis for $(\Delta J_z)^2$.

%Figura Jz & (Delta Jz)^2

\begin{figure}[h]
\scalebox{0.9}{\includegraphics{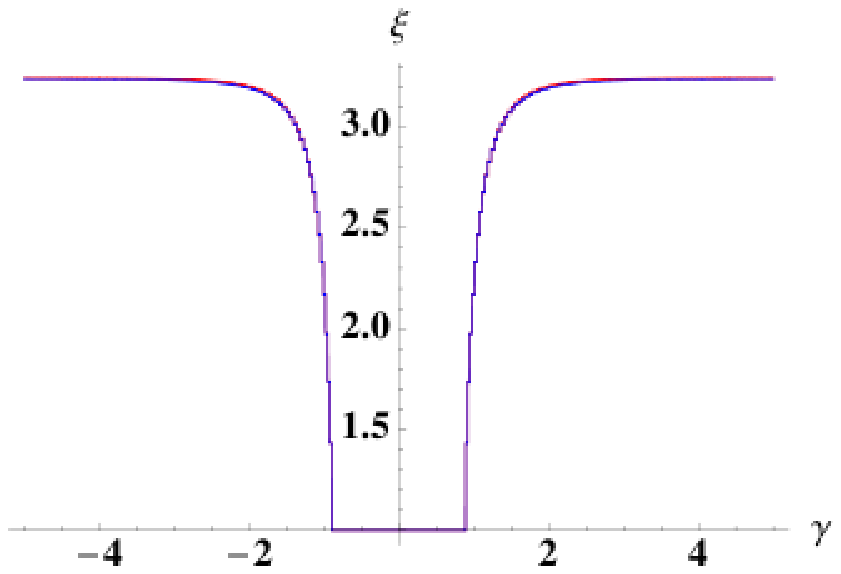}}
\qquad
\scalebox{0.9}{\includegraphics{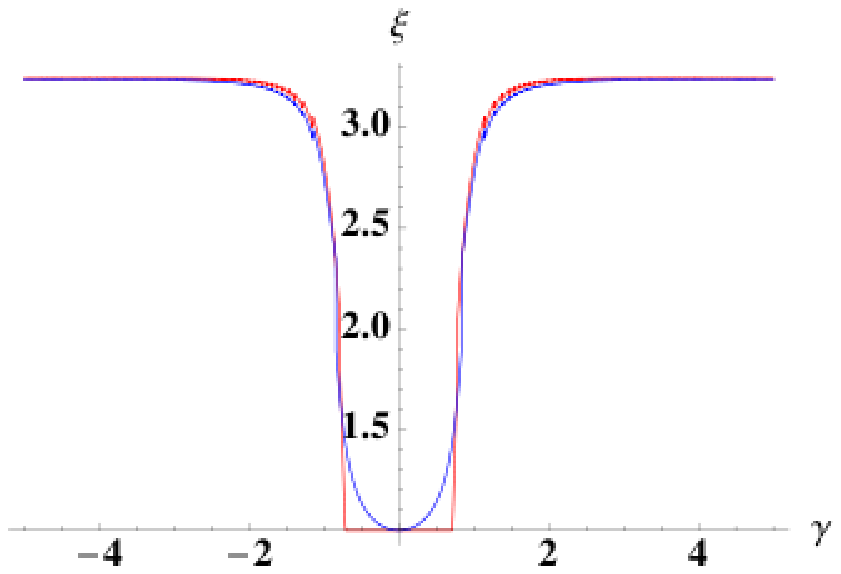}}
\caption{\label{xi}
Squeezing parameter $\xi$ of the projected state solution compared with the exact quantum solution, as functions of the interaction strength $\gamma$ for $N=20$ atoms. The plots correspond to a detuning parameter $\Delta=0.2$ (left) and $\Delta=1.5$ (right). The round shape of the pole is badly approximated at its boundary, when $\Delta > 1$, while the approximation is exact for $\Delta <1$.}
\end{figure}

As a signature of the goodness of our trial projected state in reproducing the exact quantum solution, one may use the behavior of the squeezing spin coefficient $\xi$ as given by Eq.~(\ref{squeezing}). Even though it greatly exceeds the value of $1$ away from the poles (cf. Fig.(\ref{xi})), thus suggesting a small overlap with the quantum state, both results lie exactly on top of each other.  Once again, for a large detuning making $\omega_A < 0$, the round shape of the pole is badly approximated at its boundary. For $\Delta < 1$, however, the approximation is exact.

%Figura xi

By taking the trace with respect to the field (matter) states the reduced density matrix takes the form
\begin{eqnarray}
\varrho^{matter} &=& \sum_{n=0}^{\min\{\lambda + \frac{N}{2},\,N\}} \left| c_{\lambda + \frac{N}{2} - n} \right|^2 \vert N-n,\,n\rangle \langle N-n,\,n\vert \\
\varrho^{field} &=& \sum_{\nu=\max\{0,\lambda - j\}}^{\lambda + j} \left| c_{\nu} \right|^2 \vert \nu\rangle \langle \nu\vert
\end{eqnarray}
where $c_{\lambda + \frac{N}{2} - n}$ (or $c_\nu$) is determined from the Hamiltonian diagonalization. As the reduced density matrix of the matter is diagonal, the matter-field entanglement entropy equals that between the atoms occupying the two hyperfine levels
	\begin{equation}
		S_E = - \sum_{n=0}^{\min\{\lambda + \frac{N}{2},\,N\}} \left| c_{\lambda + \frac{N}{2} - n}\right|^2 \ln 			\left| 	c_{\lambda + \frac{N}{2} - n}\right|^2
	\end{equation}
in both cases. This may be compared with that for the projected state, Ec.~({\ref{se}), to give the result shown in Fig.(\ref{Se}). The same behavior of excellent approximation for $\Delta <1$, and poor approximation at the south pole boundary when $\Delta > 1$, is obtained.

%Figura Se
\begin{figure}[h]
\scalebox{0.9}{\includegraphics{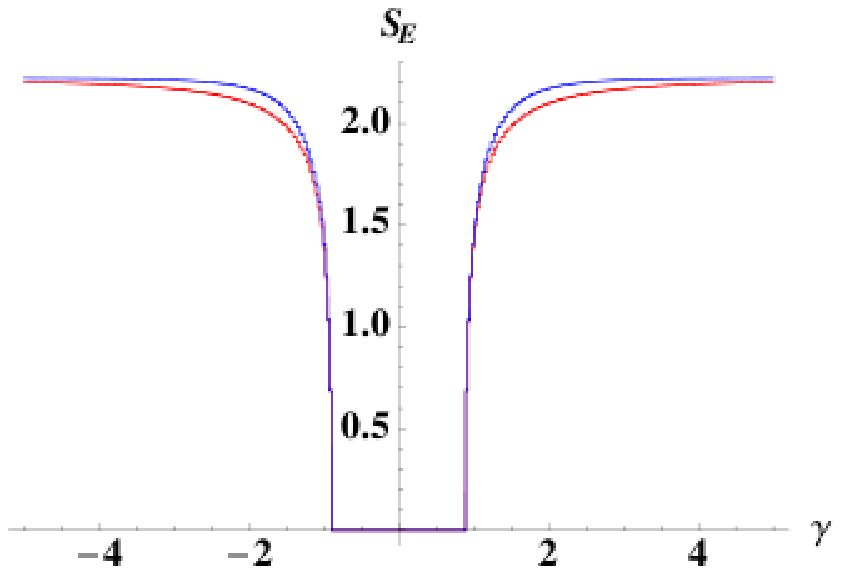}}
\qquad
\scalebox{0.9}{\includegraphics{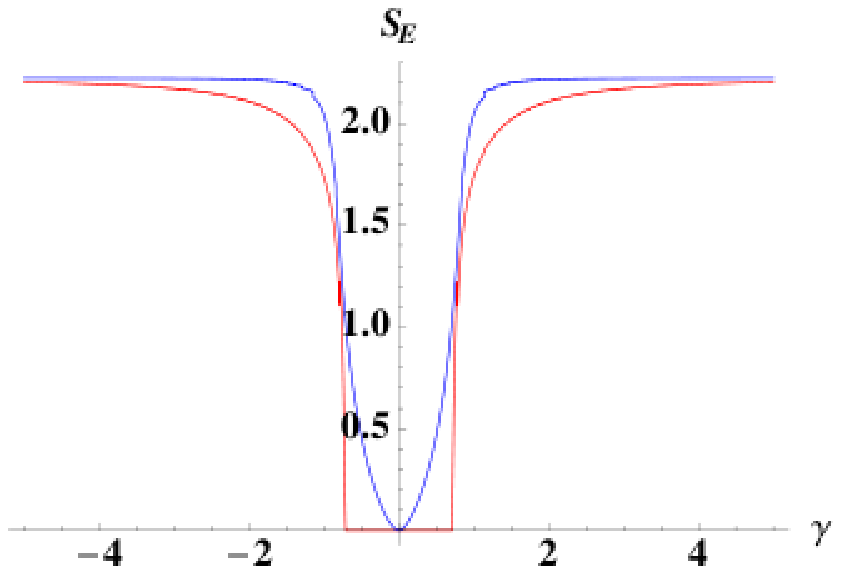}}
\caption{\label{Se}
Entanglement entropy $S_E$ of the projected state solution compared with the exact quantum solution, as functions of the interaction strength $\gamma$ for $N=20$ atoms. The plots correspond to a detuning parameter $\Delta=0.2$ (left) and $\Delta=1.5$ (right). The round shape of the pole is badly approximated at its boundary, when $\Delta > 1$, while the approximation is excellent for $\Delta <1$.}
\end{figure}

\section{Discussion and Conclusions}

Our approximation is not exact: several important observables such as $\xi,\ S_E,\ (\Delta q)^2,\ \langle \hat{n} \rangle$, and $(\Delta \hat{n})^2$ are symmetric under a reflection about $\Delta = 1$ for the projected state, while  this cannot be true for any observable coming from the Hamiltonian~(\ref{jaynes-cummings}); as an illustration of this we show $(\Delta \hat{n})^2/N^2$ as a function of $\Delta$, for both the projected state (light curve) and the quantum state (dark curve), in Fig.(\ref{fn2projq}), for $\gamma=0.75$.  While the projected state has a constant value of $0$ at both poles, the quantum state decays to $0$ asymptotically as it enters the South Pole.

%Figura Fluc n vs Delta
\begin{figure}[h]
\scalebox{1}{\includegraphics{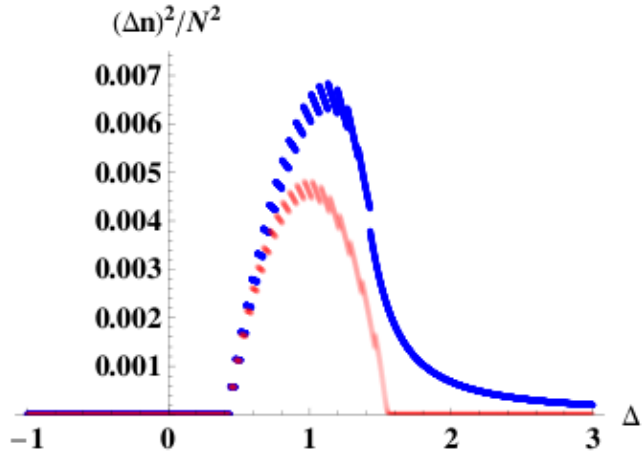}}
\caption{\label{fn2projq}
$(\Delta n)^2$ seen as a function of of $\Delta$ for the classical projected state (light curve), compared with the exact quantum solution (dark curve), for $N=20$ atoms and $\gamma=0.75$.}
\end{figure}

A good measure of the distance between quantum mechanical states is given by the fidelity; for pure quantum states it measures their distinguishability in the sense of statistical distance~\cite{wooters}, but it is customary to use the fidelity as a {\it transition probability} regardless of whether the states are pure or not.  Measuring the fidelity between our projected state and the exact quantum ground states
	\begin{equation}
		F = \vert \langle \psi_{proj} \vert \psi_{gs} \rangle \vert^2
		\label{fidelityec}
	\end{equation}
gives a result very close to 1, except in the region of classical phase transitions.  Fig.(\ref{fidelity}) (left) shows the result as a function of $\gamma$ for a detuning parameter of $\Delta = 0.2$. $F=1$ inside the North Pole and drops to $F = 0.996$ when crossing the separatrix into the Parallels region, only to approach its value of $1$ again as $\vert\gamma\vert$ continues to increase.  We have seen that the South Pole is less well represented by the projected state for $\Delta > 1$; the behavior of $F$ as a function of $\Delta$ near the South Pole is shown in Fig.(\ref{fidelity}) (right), for $\gamma = 0.75$, for which value the South Pole lies at $\Delta > 1.56$.  It is seen that even in this case $F$ drops to approximately $0.6$ and quickly recovers.

%FiguraFidelity
\begin{figure}[h]
\scalebox{0.9}{\includegraphics{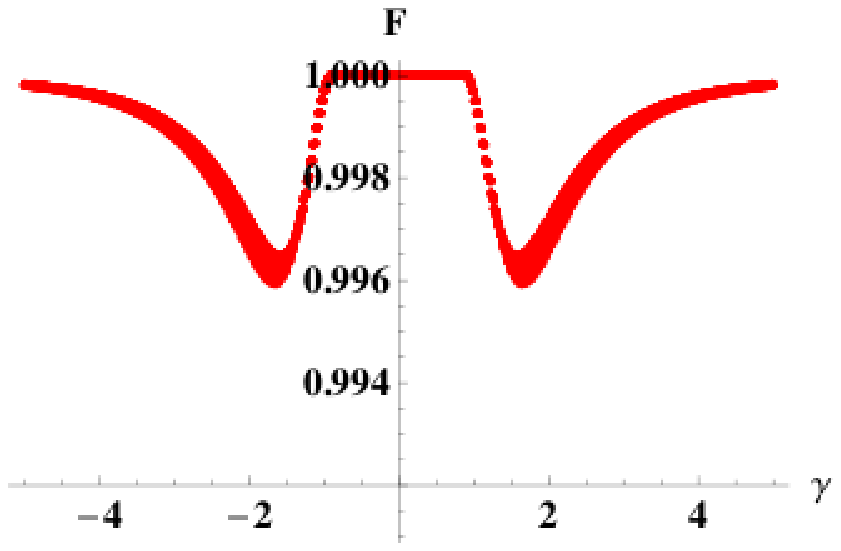}}
\qquad
\scalebox{0.9}{\includegraphics{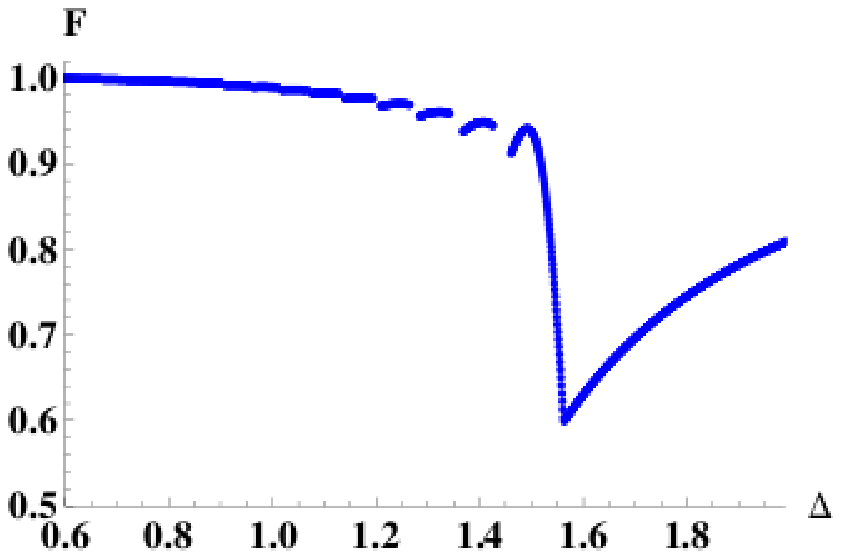}}
\caption{\label{fidelity}
Fidelity $F$ between projected state and exact quantum ground states, as a function of the interaction strength $\gamma$ (left) and of the detuning parameter $\Delta$ (right), for $N = 20$ atoms. The graph on the left corresponds to a detuning parameter $\Delta = 0.2$, and shows $F=1$ inside the North Pole dropping to $F = 0.996$ when crossing the separatrix into the Parallels region.  The one on the right corresponds to an interaction strength $\gamma = 0.75$, showing $F \approx 0.6$ at the South Pole.}
\end{figure}

It has been emphasized that the Dicke model presents quantum phase transitions when $\omega_F \gamma = \sqrt{\tilde{\omega_A}}/2$. Though difficult to satisfy for optical systems, some proposals to overcome the experimental problems have been reported~\cite{hayato}.

More recently~\cite{fink}, the Tavis-Cummings Hamiltonian has been physically realized for artificial atoms in the form of superconducting q-bits at fixed positions and coupled to a resonant cavity mode. The predicted $\sqrt{N}$-behavior for the collective $N$-atom interaction strength of the model is observed experimentally in good agreement. It is argued that the presented approach may enable novel investigations of superradiant and subradiant states of artiÞcial atoms.

In this work we have shown that an excellent approximation to the exact quantum solution of the ground state of the Tavis-Cummings model is obtained by means of a semi-classical projected state. This state has an analytical form in terms of the model parameters and allows for the analytical calculation of the expectation values of field and matter observables, entanglement entropy between field and matter, and squeezing parameter.  In our discussions we have taken the picture of two-level atoms interacting with a single mode electromagnetic field, but by re-interpreting the parameters the results are useful to describe, for example, the superradiant properties of condensed matter phases, or linear ion traps proposed for quantum computation.

\acknowledgments

This work was partially supported by CONACyT-M\'exico and DGAPA-UNAM.

\section*{Appendix A. Expectation Values of Field Operators}

The expectation value of the photon number operator can be found analytically using the overlap~(\ref{traslape}), with $\zeta^{\prime}=\zeta$, to get
	\begin{equation}
		\langle \hat{n}\rangle=\left\{\begin{array}{ll}
		(j+\lambda)-2j\,L_{j+\lambda-1}^{j-
		\lambda}(-\eta)\Big/L_{j+\lambda}^{j-\lambda}(-\eta)&
		,\quad -j+1\le\lambda\le j \\
		&\\
		(\lambda+j)-(\lambda+j)\,
		L_{2j-1}^{\lambda-j}(-\eta)\Big/
		L_{2j}^{\lambda-j}(-\eta)&,\quad\lambda\ge j
		\end{array}\right.
		\label{n-ph}
	\end{equation}
In the same way we find that the squared fluctuations are
	\begin{equation}
		\left(\Delta \hat{n}\right)^{2}=\left\{
		\begin{array}{ll}
		2j\,\Big(L_{j+\lambda-1}^{j-\lambda}(-x)\Big/
		L_{j+\lambda}^{j-\lambda}(-x)
		+(2j-1)\,L_{j+\lambda-2}^{j-\lambda}(-x)\Big/
		L_{j+\lambda}^{j-\lambda}(-x)&\\
		\quad-2j\,\left[
		L_{j+\lambda-1}^{j-\lambda}(-x)\Big/
		L_{j+\lambda}^{j-\lambda}(-x)\right]^{2}\Big)&
		,\quad -j+1\le\lambda\le j \\
		&\\
		(\lambda+j)\,\Big(L_{2j-1}^{\lambda-j}(-x)\Big/
		L_{2j}^{\lambda-j}(-x)+(\lambda+j-1)\,
		L_{2j-2}^{\lambda-j}(-x)\Big/
		L_{2j}^{\lambda-j}(-x)&\\
		\quad-(\lambda+j)\,\left[
		L_{2j-1}^{\lambda-j}(-x)\Big/
		L_{2j}^{\lambda-j}(-x)\right]^{2}\Big)&
		,\quad\lambda\ge j
		\end{array}\right.
	\end{equation}

\end{document}